\documentclass[aps,prb,twocolumn,amsmath,amssymb,superscriptaddress,floatfix]{revtex4}
\usepackage{graphicx}
\bibstyle{apsrev.bib}

\usepackage{subfigure}

\newcommand{\bc}{\begin{center}}
\newcommand{\ec}{\end{center}}
\newcommand{\be}{\begin{equation}}
\newcommand{\ee}{\end{equation}}
\newcommand{\ba}{\begin{array}}
\newcommand{\ea}{\end{array}}
\newcommand{\beq}{\begin{eqnarray}}
\newcommand{\eeq}{\end{eqnarray}}

\begin{document}

\title{Geometrical clusters in two-dimensional random-field Ising models}

\author{L\'aszl\'o K\"ornyei}
 \affiliation{Institute of Theoretical Physics,
Szeged University, H-6720 Szeged, Hungary}

\author{Ferenc Igl\'oi}
 \email{igloi@szfki.hu}
 \affiliation{Research Institute for Solid State Physics and Optics,
H-1525 Budapest, P.O.Box 49, Hungary}
 \affiliation{Institute of Theoretical Physics,
Szeged University, H-6720 Szeged, Hungary}

\date{\today}

\begin{abstract}
We consider geometrical clusters (i.e. domains of parallel spins) in the square
lattice random field Ising model by varying the strength of the Gaussian random
field, $\Delta$. In agreement with the conclusion of previous investigation
(Phys. Rev. E{\bf 63}, 066109 (2001)),
the geometrical correlation length, i.e. the average size of the clusters, $\xi$,
is finite for $\Delta > \Delta_c \approx 1.65$ and
divergent for $\Delta \le \Delta_c$. The scaling function of the distribution of the mass
of the clusters as well as the geometrical correlation function are found to involve the
scaling exponents of critical percolation. On the other hand the divergence of the
correlation length, $\xi(\Delta) \sim (\Delta - \Delta_c)^{-\nu}$, with $\nu \approx 2.$
is related to that of tricritical percolation. It is verified numerically that critical geometrical
correlations transform conformally.

\end{abstract}


\maketitle

\section{Introduction}
The random-field Ising model (RFIM) is a prototype of random systems in which the
disorder is coupled to the order-parameter of the system\cite{natt}. It has experimental
realizations, such as a diluted antiferromagnet in a field\cite{exp}. For some time there
was a debate about the lower critical dimension, $d_l$, of the systems. While
domain-wall stability arguments by Imry and Ma\cite{imry_ma} predicted $d_l=2$, perturbative field-theoretical
calculations\cite{fieldth} have lead to a different value of $d_l=3$. Later, Fisher argued\cite{fisher} that
due to the existence of several metastabile states the perturbative renormalization
does not work. Indeed, exact results by Bricmont and Kupiainen\cite{bricmont} show that in the 3d RFIM
there is ferromagnetic order and later Aizenman and Wehr\cite{aizenman} has proven rigorously that at
$d=2$ the Gibbs state is unique, thus $d_l=2$.

Although there is no ferromagnetic long-range order in the RFIM at $d=2$ there are several
interesting questions which are related to the structure of clusters
of parallel spins and the corresponding geometrical correlations in the system. Even at zero temperature,
at $T=0$, the ground state of the system is not trivial. This is the result of a competition between
the ordering effect of the ferromagnetic nearest-neighbor interaction, $J$, and the
disordering effect of the random field with a strength, $\Delta$, which is
given by the variance of the distribution. In the limit of strong random fields, $\Delta/J \gg 1$, the direction
of the spins follows the actual direction of the random fields and the ground-state is
equivalent to a site-percolation problem\cite{staufferaharony} in which an up (down) spin corresponds to an occupied (empty)
lattice site. Since the occupation probability, $p=0.5$, is below the site-percolation
threshold\cite{staufferaharony}, $p_c=0.593$, in the ground state the domains of the parallel spins have only a
finite extent, the linear size of which, $\xi$, is given by the correlation length of percolation
at $p=0.5$. As the strengths of the random field is decreased there is a tendency of the formation
of larger parallel domains, the typical size of which is a monotonously
increasing with decreasing $\Delta$. It is an interesting question, if $\xi$ stays
finite for any $\Delta>0$, or there is a finite limiting value, $\Delta_c>0$, at which
$\xi$ becomes divergent. According to recent numerical work\cite{seppala} this second scenario
holds, so that for weak enough random fields, $\Delta < \Delta_c$, the clusters of parallel
spins, which are called geometrical clusters, percolate the sample.

Geometrical clusters in non-random spin systems, such as in the Ising
and the Potts models, have been defined for a long time \cite{droplet} and their properties
have been intensively studied\cite{coniglio} at a finite temperature, $T>0$.
In two dimensions the
geometrical clusters percolate at a temperature, $T_g$, which corresponds to
the critical point of the system\cite{coniglio1}: $T_g=T_c$. In 2d their fractal dimension
can be obtained through conformal invariance\cite{2dgeom} and this value is generally
different from the fractal dimension of the so called Fortuin-Kasteleyn clusters\cite{FK}.
The Fortuin-Kasteleyn clusters are
represented by graphs of the high-temperature series expansion of the models and the corresponding
fractal dimension is directly related to the scaling dimension of the
order parameter. In three dimensions
the geometrical clusters of any spin orientations percolate in the complete
paramagnetic phase. Here the percolation transition temperature, $T_g$, is defined for the
minority spin orientation, so that for $T<T_g<T_c$ the geometrical clusters of
minority spins do not percolate\cite{3d}.
In this respect geometrical clusters play a somewhat analogous r\^ole in the 2d RFIM at $T=0$
(by varying $\Delta$) as the geometrical clusters of minority spins in the 3d Ising model for
$T<T_c$ (by varying $T$).

In this paper we consider the properties of the geometrical clusters in the 2d RFIM. In previous
numerical work\cite{seppala} a homogeneous field, $H$, was also applied and the behavior of the magnetization
and the susceptibility is studied by varying $H$ and $\Delta$. The obtained schematic phase-diagram of the
model is presented in Fig.\ref{Fig:phase}, which will be discussed in details in Sec.\ref{sec:model}.
In another work\cite{epl}
nonequilibrium critical relaxation of the 2d Ising model has been studied, in which the
initial state was prepared as the ground state of the RFIM. By varying $\Delta$ and setting
$H=0$ the geometrical phase transition is found to influence the properties of
the nonequilibrium dynamical processes by introducing new finite time and length scales.

In the present paper we study
those aspects of the geometrical clusters in the RFIM which have not been yet investigated in previous work.
Here we set
$H=0$, vary the strength of disorder and study the distribution of the mass, $M$ of the geometrical clusters,
$R(M,L)$ in a finite system of linear size, $L$. In particular from the scaling form of $R(M,L)$
we calculate both the fractal dimension, $d_f$,
and the distribution exponent\cite{staufferaharony}, $\tau$. We also
measure the geometrical correlations, $G(r)$. This is the average over all spin pairs of distance, $r$,
with a contribution one, if both spins are in the same geometrical cluster and zero otherwise. We show that $G(r)$ plays the role of a standard correlation function in the vicinity of
a second-order phase transition point. The correlation length associated
to $G(r)$, $\xi(\Delta)$, is
found to be finite for $\Delta > \Delta_c$ and divergent for $\Delta \le \Delta_c$, thus defining
two different regions, see in Fig.\ref{Fig:phase}. In the percolating
regime, $\Delta \le \Delta_c$, $G(r)$ exhibits quasi-long-range order, $G(r) \sim r^{-\eta}$. We have also studied conformal aspects of $G_r$ in the percolating regime.

The structure of the paper is the following. The model, the numerical method and the
known results are given in Sec.\ref{sec:model}. Our results are presented in Sec.\ref{sec:results} and
discussed in Sec.\ref{sec:disc}.

\section{Model, numerical method and known results}
\label{sec:model}

The RFIM is defined by the Hamiltonian:
\be
{\cal H}=-J \sum_{\langle ij \rangle} \sigma_i \sigma_j -\sum_i (h_i+H) \sigma_i\;,
\label{HRFIM}
\ee
where $\sigma_i=\pm 1$ is an Ising spin located at site $i$ of a square lattice. The ferromagnetic
nearest-neighbor coupling is chosen to be $J=1$, whereas the $h_i$ random
fields are independent variables which are taken from a symmetric Gaussian distribution:
\be
P(h_i)=\frac{1}{\sqrt{2 \pi \Delta^2}} \exp\left[-\frac{h_i^2}{2 \Delta^2} \right]\;,
\label{PG}
\ee
which has a variance $\Delta$. For completeness in Eq.(\ref{PG}) we have also included a
homogeneous field of strength, $H$. During our numerical calculations we shall put $H=0$.

We are interested in the properties of the system at $T=0$ when all information is
encoded in its ground state. The problem of finding the ground state of the RFIM
is a non-trivial task, but can be exactly solved through a mapping to a maximum flow
problem\cite{jc}.
For this latter problem there are very efficient combinatorial optimization
algorithms\cite{heiko} which work in strongly polynomial time. With this method finite systems as
large as $L=500$ can be treated numerically.

At $H=0$ and for $\Delta>0$ in the ground state of the RFIM there is no magnetic long-range order\cite{aizenman}.
This can be illustrated by the average spin-spin correlation function,
$C(r)=\overline{\langle \sigma_0 \sigma_r \rangle}$, which is averaged over all pairs of spins
having a distance $r$ apart and $\langle \dots \rangle$ stands for the calculation in the
ground state. $C(r)$ goes to zero exponentially with the distance, $r$. In the presence of a homogeneous field,
$H \ne 0$, there is a finite magnetization, $m(H)=\overline{\langle \sigma \rangle} \ne 0$,
and one can measure the magnetic susceptibility, $\chi(H)$. By varying $H$, however, $\chi(H)$
displays no singularity\cite{seppala}.

The ground state of the RFIM can be visualized by denoting the up and down spins by
different symbols. In this picture the domains of parallel spins form
geometrical clusters, the structure of which depends on the value of $\Delta$ (and also $H$).
The typical size of the clusters (of majority spins), $\xi$, serves as a bulk
length-scale in the problem. Between large, oppositely magnetized clusters there is an interface,
which is smooth for small lengths, up to $l<l_b$, and its
roughness is seen only for larger scales, for $l>l_b$. Here $l_b$ is the
so called breaking-up length. For weak random
fields the breaking-up length asymptotically behaves as\cite{lb}
\be
l_b \sim \exp(A/\Delta^2)\;,
\label{l_b}
\ee
thus it is divergent as $\Delta \to 0$. The existence of the breaking-up length
imposes limitations on the possible numerical calculations. The linear size
of the system, $L$, should be much larger than $l_b$, therefore one can not use
a too small value of $\Delta$. Due to this reason we have not reduced $\Delta$
below $1.2-1.4$ for the system sizes we could treat numerically.

\begin{figure}[h!]
  \begin{center}
     \includegraphics[width=2.5in,angle=0]{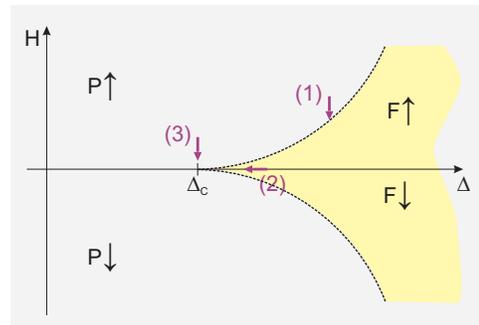}
   \end{center}
   \caption{
(Color online) Schematic phase-diagram of geometrical clusters of the 2d RFIM as obtained in
Ref.[\onlinecite{seppala}]. In the regions, $F\uparrow$ and $F\downarrow$, the
majority spin clusters are non-percolating, in $P\uparrow$ and $P\downarrow$, they
are percolating. The phase boundary is indicated by dashed lines.
The transition along arrow (1) is in the standard percolation
universality class. Investigations in this paper are restricted to $H=0$, in
particular in the vicinity of the multicritical point along arrow (2).
}
   \label{Fig:phase}
 \end{figure}

Some aspects of the structure of the geometrical clusters for a non-zero homogeneous field has been studied in Ref.[\onlinecite{seppala}], see the phase-diagram in Fig.\ref{Fig:phase}.
For a fixed $\Delta>\Delta_c$ the size of the majority spin clusters, $\xi$, is finite
for small $H$, but monotonously
increasing with increasing $H$. There is a threshold value, $H_p(\Delta)$, when the majority spins
percolate the sample and for $H>H_p(\Delta)$ a finite fraction of the majority spins belongs to
the infinite cluster. Close to the percolation point, see arrow $(1)$ in Fig.\ref{Fig:phase}, the percolation probability is found to
depend on the scaling combination: $L(H_p-H)^{\nu_p}$, with $\nu_p=4/3$ characteristic
of the 2d short range percolation transition point\cite{staufferaharony}. Also the fractal dimension of the
geometrical clusters at the percolation point is found\cite{seppala} to be compatible with that of
standard percolation\cite{staufferaharony}, $d_f=91/48$. According to the numerical results
as $\Delta$ is decreased $H_p$ tends to zero at $\Delta_c$, as $H_p \sim (\Delta-\Delta_c)^{\phi}$,
with $\Delta_c \approx 1.65$ and $\phi \approx 2.05$.

In the following section we concentrate on the properties of the geometrical
clusters at $H=0$. In particular we study the distribution of their mass and
introduce and investigate a correlation function, which is associated to geometrical clusters.
In these investigations we approach the multicritical point at $\Delta=\Delta_c$ and $H=0$
along arrow $(2)$ in Fig.\ref{Fig:phase}.
With these investigations we want to shied light to
the possible geometrical phase transition of the system which takes place at
$\Delta_c$.

\section{Results}
\label{sec:results}
We have considered the RFIM on $L \times L$ square lattices with periodic boundary
conditions in both directions. The strength
of the random field is varied in the range $1.2 < \Delta < 4.$, where the lower limiting value is
set by the condition, $L \gg l_b(\Delta)$, where the breaking-up length is given
in Eq.(\ref{l_b}). Using combinatorial optimization algorithm we have calculated
the ground state configuration exactly for different finite samples up to a linear size $L=500$.
Averages are performed over typically $10000$ different realizations
of disorder.

\subsection{Distribution of the mass of geometrical clusters}

\begin{figure}[h!]
  \begin{center}
     \includegraphics[width=2.5in,angle=0]{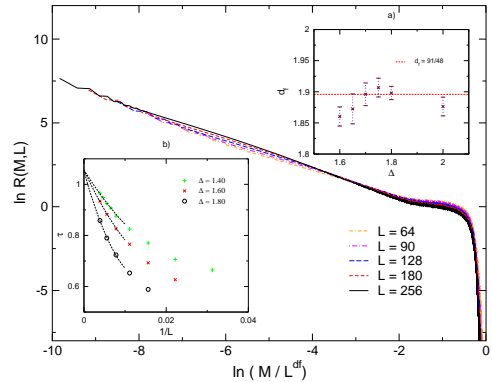}
   \end{center}
   \caption{
(Color online) Scaling plot of the cumulative distribution of the mass of the geometrical clusters at $\Delta=1.8$, using the fractal dimension of standard percolation. In the inset a) the effective fractal dimension is presented as calculated from the optimal scaling collapse of the curves for different values of $\Delta$. In the
inset b) extrapolation of the $\tau$ exponents calculated at different finite systems is presented
for $\Delta=1.4$ ($\times$), $\Delta=1.6$ ($+$) and $\Delta=1.8$ ($*$).
}
   \label{Fig:mass}
 \end{figure}

We start to study the cumulative distribution of the mass (number of spins), $M$, of the clusters,
which is denoted by $R(M,L)$. This measures
the fraction of clusters having at least a mass of $M$. According to scaling
theory developed for standard percolation\cite{staufferaharony} this distribution asymptotically
behaves as:
\begin{equation}
R(M,L)=M^{-\tau} \tilde{R}(M/L^{d_f})
\label{eq:R_mL}
\end{equation}
with a characteristic exponent, $\tau=2/d_f$. Thus the only parameter which enters into this
expression is the fractal dimension of the clusters, $d_f$. This relation is expected to hold at the critical point, where the correlation length, $\xi$, associated to geometrical correlations is divergent.
However, Eq.(\ref{eq:R_mL}) can be a good approximation outside the critical point, too,
provided the correlation length, $\xi$, is much larger than the size of the system, $L$. To compute
$R(M,L)$ we have performed cluster statistics over $10000$ different random samples and the distributions
obtained for different finite systems are scaled together using the relation in Eq.(\ref{eq:R_mL}).
As an illustration in Fig.\ref{Fig:mass} we present the scaled cumulative distribution functions at
the specific value, $\Delta=1.8$, in which we used the fractal dimension of
standard percolation\cite{staufferaharony}, $d_f=91/48$.
The scaling collapse in Fig.\ref{Fig:mass} is indeed satisfactory.
Repeating the same procedure in the range of $1.4 < \Delta < 2.6$ we have obtained still acceptable data collapse. This means that in this regime the condition $\xi \gg L$ is satisfied and we can
conclude that the critical point, $\Delta_c$, is also located in this range. 

We have also estimated the value of the fractal dimension from the optimal scaling collapse of the curves.
For this we have measured the surface of the overlap of the scaled curves for different values of the
parameter, $d_f$, and at the true fractal dimension the overlap surface is minimal, see in Ref.\cite{long2d}.
The measured effective, i.e. $\Delta$-dependent
fractal dimensions are presented in the inset a) of Fig.\ref{Fig:mass}. In the range of
$1.6 < \Delta < 2.$ the effective fractal dimensions are approximately constant and
consistent with $d_f=1.89(2)$. For $\Delta>2.$ the effective $d_f$ start to decrease, which is a clear indication that the correlation length is not large enough. The decrease of $d_f$ for $\Delta<1.6$,
is probably due to the large value of the breaking-up lengths, which starts
to approach the size of the system.

From the slope of the curves in Fig.\ref{Fig:mass} in the log-log plot we have measured the exponent, $\tau$,
which is defined in Eq.(\ref{eq:R_mL}). The measured effective exponents for different finite sizes
are plotted in the inset b) of Fig.\ref{Fig:mass} for three different values of $\Delta$. Here 
strong finite-size dependence is visible so that we have to perform an extrapolation procedure
in terms of $1/L$. The extrapolated value, as seen in the inset b) of Fig.\ref{Fig:mass},
is given by $\tau=1.055(3)$, and this is independent of $\Delta$ we considered. We note that
our estimate fits very well to the theoretical value of standard percolation\cite{staufferaharony}: $\tau=96/91=1.055$. We can thus conclude that the distribution function of the mass of the geometrical
clusters nicely satisfies the scaling prediction in Eq.(\ref{eq:R_mL}) and the fractal dimension
coincides with that of standard percolation.

\subsection{Geometrical correlations}

Here - using the analogy with ordinary percolation - we introduce the geometrical correlation
function, $G(r)$, which is associated to the
geometrical clusters. By definition between two spins the geometrical correlation is one,
if both spins belong to the same
geometrical cluster and zero otherwise. $G(r)$ is then obtained by averaging the correlations
over all pairs of spins having a distance, $r$. In the non-critical state, where $\xi < \infty$,
the geometrical correlation function is short ranged and expected to decay asymptotically
as $G(r) \sim \exp(-r/\xi)$.
On the other hand at the critical point the geometrical correlation function has an
algebraic decay, $G(r) \sim r^{-\eta}$, where the decay exponent, $\eta$, is related
to the fractal dimension as, $\eta=2(d-d_f)$, through scaling theory.

\begin{figure}[h!]
  \begin{center}
     \includegraphics[width=2.5in,angle=0]{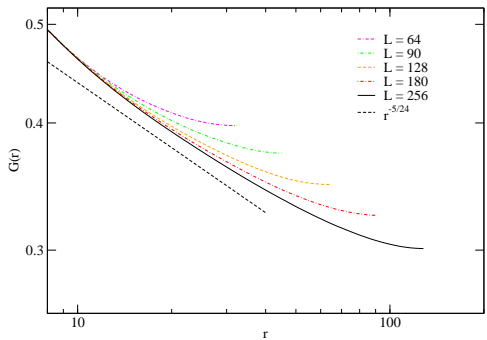}
   \end{center}
   \caption{
(Color online) Geometrical correlations in a log-log plot for $\Delta=1.8$ and for different
system sizes. The theoretical prediction of the asymptotic behavior is indicated
by a straight dotted line.
}
   \label{Fig:corr}
 \end{figure}

\begin{figure}[h!]
  \begin{center}
     \includegraphics[width=2.5in,angle=0]{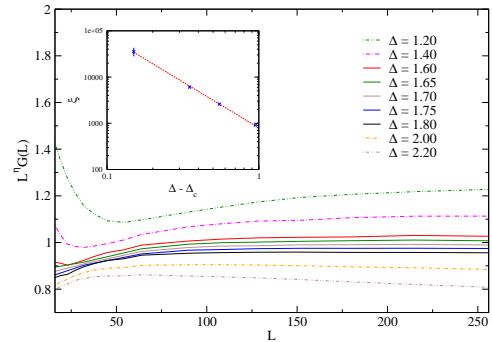}
   \end{center}
   \caption{
(Color online) Scaled maximal distance geometrical correlations for different values of $\Delta$.
The scaled correlations seem to reach a constant limiting value for $\Delta \le 1.6$, thus
here the correlation length is divergent. For stronger random fields the correlations decay to zero,
although the correlation length is still a large finite value. Inset: the correlation length as
a function of $\Delta-\Delta_c$ in a log-log plot. The best fit is obtained with $\Delta_c=1.65$
and with an exponent, $\nu=1.98(5)$.
}
   \label{Fig:corrL}
 \end{figure}

In practical calculations the system has a finite size, $L$, and in the critical region
the geometrical correlation function, $G(r,L)$, is expected to behave asymptotically as: $G(r,L)=r^{-\eta} \tilde{G}(r/L)$. Here the scaling function, $\tilde{G}(y)$, should approach a limiting
finite value for small arguments, so that
the decay exponent, $\eta$, can be calculated for large finite systems, if $1 \ll r \ll L$.
As an illustration we shown in
Fig.\ref{Fig:corr} the geometrical correlation function for different finite
systems in log-log plot for $\Delta=1.8$. Indeed the slope of the curves (of the largest systems)
in the region $10<r<40$
is compatible with the theoretical prediction of ordinary percolation: $\eta=5/24=0.208$. 
\vskip 1cm
In order to obtain a more accurate estimate of the asymptotic behavior of the geometrical correlation function
we measured the correlations between two points of maximal distance, i.e. at $r=L/2$. The maximal distance
correlation function defined in this way, $G(L)\equiv G(r=L/2,L)$, is expected to behave in
the critical region as: $G(L)=L^{-\eta} \hat{G}(L/\xi)$, thus it is advisable to consider the
scaled maximal distance correlation function, $L^{\eta} G(L)=\hat{G}(L/\xi)$. In Fig.\ref{Fig:corrL}
we have plotted this function for different values of $\Delta$.
\bigskip

Here one can identify two different regimes as far as the large $L$ dependence of the scaled
correlation function is concerned. For strong random fields, $\Delta > \Delta_c$, the scaled
correlations start to decay to zero, so that the correlation length is finite and the
geometrical clusters are non-percolating. On the other hand for weak random fields, $\Delta \le \Delta_c$,
the scaled correlations approach a finite limiting value, so that the correlation length is
divergent and we are in the phase in which the geometrical clusters are percolating. This
conclusion is the same as that obtained through the analysis of the
spanning probability in Ref.\cite{seppala}.

The threshold value of $\Delta_c$ can be calculated by analysing the behavior of the correlation
length, which can be deduced from the asymptotic dependence: $\hat{G}(L/\xi) \sim \exp[-L/(2\xi)]$.
Close to the critical point the correlation length is
expected to be in the form: $\xi(\Delta) \sim (\Delta - \Delta_c)^{-\nu}$. As illustrated in the inset
of Fig.\ref{Fig:corrL} the
best fit is obtained with $\Delta_c \approx 1.65$, as in Ref.\cite{seppala}, whereas for the
correlation length critical exponent we have obtained: $\nu=1.98(5)$. This exponent describes
the singularity of $\xi$ along arrow (2) in Fig.\ref{Fig:phase}. If we want to calculate
the divergence of the correlation length at $\Delta_c$ but with a small homogeneous field, $H$, i.e.
along arrow (3) in Fig.\ref{Fig:phase}, we
make use of the known result about the critical boundary, $H_p \sim (\Delta-\Delta_c)^{\phi}$. Than
from scaling theory we obtain $\xi(H_p,\Delta_c) \sim \xi(0,\Delta)\sim H_p^{-\tilde{\nu}}$,
with $\tilde{\nu}=\nu/\phi=0.97(5)$. We note that this value corresponds to the inverse of
the second thermal exponent of tricritical percolation\cite{nienhuis}: $1/y_{t2}=1$.

\subsection{Geometrical correlations in the strip geometry}

\begin{figure}[h!]
  \begin{center}
     \includegraphics[width=2.5in,angle=0]{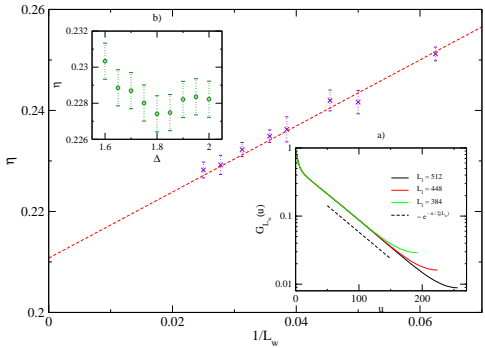}
   \end{center}
   \caption{
(Color online) Effective decay exponents calculated in periodic strips of width $L_w$
through Eq.(\ref{corr_exp}). The dashed straight line indicates extrapolation
through $1/L_w$. Inset a): Geometrical correlation function at $\Delta=1.8$ in a strip of
width $L_w=40$ and for different lengths, $L_1$. In a semi-log plot the slope of
the limiting curve is indicated by a dashed line and given by $1/\xi(L_w)$.
Inset b): $\Delta$-dependence of the effective
decay exponent for $L_w=40$.
}
   \label{Fig:strip}
 \end{figure}

Critical correlations are generally invariant under conformal transformations
and conformal invariance has very important consequences in two dimensional systems.\cite{cardy}
One important application of the conformal method is to transform the correlation functions
from one geometry to another. In this respect it is of interest if also
geometrical correlations are conformally invariant and if one can transform them between different
geometries. To answer to this question we have studied geometrical correlations
in the RFIM also in the strip geometry. It is known that the infinite plane, described by the
complex variable, $z=x+iy$, can be mapped into a periodic strip of width, $L_w$, having the
complex variable, $w=u+iv$, through the logarithmic conformal transformation:
\be
w=\frac{L_w}{2\pi} \ln z\;.
\label{log_map}
\ee
Then critical correlations in the plane, $G(z) \sim |z|^{-\eta}$, are transformed into
an asymptotic exponential decay along the strip:
\be
G_{L_w}(u) \sim \exp(-u/\xi(L_w))\;,
\label{corr_strip}
\ee
where the correlation length, $\xi(L_w)$, is proportional to the width of the strip
and given by:
\be
\xi(L_w)=\frac{L_w}{\pi \eta}\;.
\label{corr_exp}
\ee
Thus the proportionality factor contains the decay exponent in the plane, therefore Eq.(\ref{corr_exp})
is called the correlation length - exponent relation\cite{luck,cardy84}.

To check the validity of the correlation length - exponent relation for the RFIM
we used strips of widths, $L_w=16,20,\dots,40$ and calculated the geometrical correlation
function along the strip. The lengths of the strips are taken so large, that the calculated
exponential decay in Eq.(\ref{corr_strip}) becomes independent of it. This is illustrated in the
inset a) of Fig.\ref{Fig:strip}.
Generally we went at least up to a length of $16\times L_w$ sites. After measuring the correlation
length for a given $L_w$ we have calculated from Eq.(\ref{corr_exp}) an effective, i.e. $L_w$-dependent
exponent. This calculation is then repeated for several values of $\Delta$ in the range of
$1.4<\Delta<2.0$. For a given $L_w$ the $\Delta$-dependence of the exponents are found
to be smaller than the actual error of the calculation, as shown in the
inset b) of Fig.\ref{Fig:strip}. Therefore in Fig.\ref{Fig:strip}
we have plotted an average, i.e. $\Delta$-independent exponent as a function of $1/L_w$. As seen in this
figure the effective exponents have a size dependence and an
extrapolation through $1/L_w$ leads to the estimate: $\eta=0.210(4)$. This value is in good agreement with the
conformal prediction: $\eta=5/24=0.208$. Thus we conclude that our numerical study has confirmed 
that the correlation length - exponent relation is valid for geometrical correlations in the RFIM.

\section{Discussion}
\label{sec:disc}

In this paper we have considered the RFIM in the square lattice and studied the structure of
its ground state at $T=0$. Since in the ground state of the system there is no ferromagnetic
long-range order we have focused on the properties of the geometrical clusters consisting of
parallel spins. In particular we have investigated the distribution of the mass of the
geometrical clusters and studied geometrical correlations, which are associated to
geometrical clusters. In agreement with previous investigations\cite{seppala} the
geometrical clusters are found to have a finite extent for strong random fields,
$\Delta > \Delta_c$, and being percolating for weak random fields, $\Delta < \Delta_c$.
For $\Delta \le \Delta_c$ the fractal dimension of geometrical clusters is calculated
by different methods and its value is found to be in good agreement with short range percolation.
The geometrical correlation function of the system is calculated both in the plane and in the
strip geometries. It is shown that the critical geometrical correlation function satisfies
both scaling and conformal invariance. In the vicinity of the geometrical critical
point the correlation length is found to diverge as: $\xi(\Delta)
\sim (\Delta-\Delta_c)^{-\nu}$, with an exponent, $\nu \approx 2$, whereas at the critical
disorder, $\Delta=\Delta_c$, the divergence of $\xi(H)$ in the presence of a homogeneous field involves the
exponent, $\tilde{\nu} \approx 1$. This latter is characteristic for {\it tricritical} percolation.

We note that our study of geometrical clusters in the RFIM is somewhat
related to the properties of the random bond Potts model (RBPM) in the large-$q$ limit. As was
shown by Cardy and Jacobsen\cite{pottstm} the interface Hamiltonian of
the RFIM separating different types of geometrical clusters, and the interface Hamiltonian of
the RBPM separating different types of Fortuin-Kasteleyn clusters, can be mapped
into each other. In this respect the absence of long-range order in the RFIM corresponds
to the absence of phase coexistence in the RBPM, thus no first-order phase transition
in the presence of bond disorder in 2d\cite{aizenman}. Although the interface Hamiltonian
of the two problems are equivalent the cluster structure, in particular the value of the
fractal dimension is different. For the RBPM the conjectured value\cite{ai03,long2d} for the Fortuin-Kasteleyn
clusters, $d_f^P=(5+\sqrt{5})/4=1.809$, is considerably smaller than that of the RFIM. Thus
arguments based on the interface Hamiltonian can be used to study the stability of some
phases, but the actual cluster structure is governed by such critical fluctuations which
are encoded in the details of the Hamiltonian.

\begin{acknowledgments}
We are indebted to M. Pleimling for cooperation in related problems and for useful discussions.
This work has been
supported by a German-Hungarian exchange program (DAAD-M\"OB) and by the
Hungarian National Research Fund under grant No OTKA TO37323,
TO48721, K62588, MO45596 and M36803.
\end{acknowledgments}

\end{document}